\documentclass{article}%
\usepackage{amsmath}
\usepackage{amsfonts}
\usepackage{amssymb}
\usepackage{graphicx}%
\newcommand{\bpartial}{\mathop{\partial\kern -4pt\raisebox{.8pt}{$|$}}}
\newcommand{\bra}{\mathopen{[\kern-1.6pt[}}
\newcommand{\ket}{\mathclose{]\kern-1.5pt]}}
\newcommand{\bbra}{\mathopen{[\kern-2.2pt[\kern-2.3pt[}}
\newcommand{\bket}{\mathclose{]\kern-2.1pt]\kern-2.3pt]}}

\makeindex
\begin{document}

\title {\LARGE {Jacobi-Lie symmetry and Jacobi-Lie T-dual sigma models on group manifolds}}
\vspace{3mm}

\author {  A. Rezaei-Aghdam \hspace{-2mm}{ \footnote{ rezaei-a@azaruniv.edu }} \hspace{2mm}{\small and}\hspace{2mm}
 M. Sephid{ \footnote{s.sephid@azaruniv.edu }}\hspace{2mm}\\
{\small{\em Department of Physics, Faculty of Sciences, Azarbaijan Shahid
}}\\
{\small{\em Madani University  , 53714-161, Tabriz, Iran  }}}

\maketitle

\vspace{3cm}

\begin{abstract}
Using the concept of Jacobi-Lie group and Jacobi-Lie bialgebra, we generalize the definition of Poisson-Lie symmetry to Jacobi-Lie symmetry. In this regard, we generalize the concept of Poisson-Lie T-duality to Jacobi-Lie T-duality and present Jacobi-Lie T-dual sigma models on Lie groups, which have Jacobi-Lie symmetry. Using this symmetry, new cases of duality appear and some examples are given. This generalization may provide insights to understand the quantum features of Poisson-Lie T-duality, in a more satisfactory way.
\end{abstract}

\newpage
\section{\bf Introduction}

Target-space duality is one of the most important symmetries of string theories and (in general) sigma models. Before introducing the Poisson-Lie symmetry \cite{KL, K} there was a common belief among the people that the existence of T-duality mainly depends on the isometry of target manifold. Depending on the kind of isometry, T-duality was called Abelian or non-Abelian (for a review see \cite{GPR, AAL}). It must be mentioned that, in non-Abelian case, the dual model may not have isometry, so, it is not possible to obtain the original model from the dual one. The requirement of isometry in acquiring the dual model was removed by Klim\v{c}ik and \v{S}evera in their pioneering work, i.e., Poisson-Lie T-duality and Poisson-Lie symmetry \cite{KL, K}. Poisson-Lie symmetry does not require the isometry in the original target manifold. In this case, the Noether's currents associated with the action of Lie group $G$ on target manifold are not conserved but the components of the Noether's currents satisfy in the flat connection condition, i.e. they satisfy the Maurer-Cartan equation with structure constants of ${\tilde{\bf g}}$ (Lie algebra of the Lie group $\tilde{G}$ with the same dimension as $G$) \cite{KL, K} such that $G$ and $\tilde{G}$ are Poisson-Lie groups \cite{Drin2} and their Lie algebras form a Lie bialgebra $({\bf g},\bf{\tilde{g}})$ \cite{Drin2, AM}. 
Up to now, several attempts in the construction of Poisson-Lie T-dual sigma models \cite{S4, JR} and investigation of Poisson-Lie symmetry of some WZW models \cite{AKT1, ER4} have been performed. The generalization of Poisson-Lie T-dual sigma models on supermanifolds has been done in \cite{ER1, ER7, ER8}. Also, canonical equivalence of the Poisson-Lie T-dual sigma models already has been shown in \cite{KL, K, KS2, S3}, but in general the quantization of these T-dual sigma models is still a challenging problem (the quantum equivalence of the Poisson-Lie T-dual sigma models under one-loop renormalization group flow is studied for  some special examples in \cite{S4, AKT1, BFMPS} and has been shown in general in \cite{SS1} and \cite{SS2}). Here, we will try to generalize the concept of Poisson-Lie symmetry and Poisson-Lie T-duality to Jacobi-Lie symmetry and Jacobi-Lie T-duality. We hope that this generalization and some new cases of this duality lead us to understand the quantum features of Poisson-Lie T-duality.

In Jacobi-Lie symmetry, we investigate the two-dimensional sigma models on target manifold alongside with $\sigma$-functions. Moreover, the components of Noether's currents satisfy in generalized flat connection condition, that is, they satisfy the generalized Maurer-Cartan equation with 1-cocycle $X_{0}$ and structure constants of the Lie algebra ${\tilde{\bf g}}$ (related to the Lie group $\tilde{G}$); such that $G$ and $\tilde{G}$ are Jacobi-Lie groups \cite{Iglesias} and their Lie algebras form a Jacobi-Lie bialgebra \cite{Iglesias, RS4}.

The outline of the paper is as follows: In second section, we define the concept of Jacobi-Lie symmetry using the ansatz that the components of Noether's currents are related to the original sigma model with target manifold $M$ (such that the model is invariant under the action of Lie group $G$ on target manifold $M$) satisfy the generalized flat connection condition. We have shown that the integrability of Jacobi-Lie symmetry gives some conditions of Jacobi-Lie bialgebras $(({\bf{g}},\phi_{0}),({\tilde{\bf g}},X_{0}))$ \cite{RS4}. Also, the integrability condition of Jacobi-Lie symmetry on the dual model has led to other conditions of Jacobi-Lie bialgebras. Furthermore, it has been shown that the Poisson-Lie symmetry can be obtained as a special case of Jacobi-Lie symmetry and also we will have other new cases. In the third section, we provide the formalism of construction of Jacobi-Lie T-dual sigma models on Lie groups. The Poisson-Lie T-dual sigma models can be obtained as a special case; furthermore, we provide some new dual models. An example of such Jacobi-Lie T-dual models is given in section four. Some discussions are given in concluding remarks.
\section{\bf Jacobi-Lie symmetry}

Here, we will try to extend the Poisson-Lie symmetry \cite{KL,K} to Jacobi-Lie symmetry. We hope that this generalization has some new insights to understand the quantum features of Poisson-Lie T-duality. We study the two-dimensional sigma model on target space $M$ with background matrix ${\cal E}_{\mu\nu}=G_{\mu\nu}+B_{\mu\nu}${\footnote{$G_{\mu\nu}$ and $B_{\mu\nu}$ are components of the metric and antisymmetric tensor field on the target manifold $M$.}} and a $\sigma$-function on $M$ ($\sigma \in C^{\infty}(M)$), as follows

\begin{equation}\label{1}
S=\frac{1}{2}\int_{\Sigma}{d\xi^{+}\wedge d\xi^{-}~e^{\sigma(x)}{\cal E}_{\mu\nu}(x)~ \partial_{+}x^{\mu}\partial_{-}x^{\nu}}=\frac{1}{2}\int_{\Sigma}{d\xi^{+}\wedge d\xi^{-}~L},
\end{equation}
where $\xi^{\pm}$ and $\{x^{\mu}\}$ are coordinates of the worldsheet $\Sigma$ and target manifold $M$, respectively. Suppose that a Lie group $G$ acts freely on manifold $M$. One can consider this action by the following transformation

\begin{equation}\label{2}
x^{\mu}\rightarrow x^{\mu}+\epsilon^{i}(\xi^{+},\xi^{-})v_{i}\hspace{0cm}^{\mu},
\end{equation}
where the left-invariant vector fields $v_{i}\hspace{0cm}^{\mu}$ corresponds to the right action of the Lie group $G$ on the target manifold $M$ and $\epsilon^{i}(\xi^{+},\xi^{-})$ is an infinitesimal function of light cone coordinate of $\Sigma$. Now, one can calculate the variation of $S$ under this transformation as follows

\begin{equation}\label{3}
\delta S=\frac{1}{2}\int{d\xi^{+}\wedge d\xi^{-}~e^{\sigma(x)}\epsilon^{i}({\cal L}_{v_{i}}}{\cal E}_{\mu\nu}+v_{i}\hspace{0cm}^{\lambda}\partial_{\lambda}\sigma~{\cal E}_{\mu\nu})~\partial_{+}x^{\mu}\partial_{-}x^{\nu}-\frac{1}{2}\int{d\epsilon^{i}\wedge \star J_{i}},
\end{equation}
where Hodge star of the Noether's currents $J_{i}$ has the following form

\begin{equation}\label{4}
\star J_{i}=e^{\sigma({x})}({\cal E}_{\mu\gamma}~v_{i}\hspace{0cm}^{\gamma}~ \partial_{+}x^{\mu}d\xi^{+}-{\cal E}_{\gamma\nu}~v_{i}\hspace{0cm}^{\gamma}~ \partial_{-}x^{\nu}d\xi^{-}).
\end{equation}
If we consider $\phi_{0}=d \sigma \in \Omega^{1}(M)$ as an 1-cocycle on ${\bf g}$ (Lie algebra of the Lie group $G$) with values in ${\tilde{\bf g}}$ {\footnote {Here, one can represent the elements of ${\tilde{\bf g}}$ (dual space of ${\bf g}$) in terms of the elements of the $1$-forms on $M$, i.e. $\Omega^{1}(M)$.}} then we have $\phi_{0}$-Lie derivative in the sense of D. Iglesias and J. C. Marrero as follows \cite{IM}

\begin{equation}\label{5}
({\cal L}_{\phi_{0}})_{v_{i}}{\cal E}_{\mu\nu}={\cal L}_{v_{i}}{\cal E}_{\mu\nu}+<\phi_{0},v_{i}>{\cal E}_{\mu\nu},
\end{equation}
where $<,>$ is the inner product between forms and vector fields on $M$. Therefore, one can rewrite the variation of $S$ using the above definition for $\phi_{0}$-Lie derivative as follows

\begin{equation}\label{6}
\delta S=\frac{1}{2}\int{d\xi^{+}\wedge d\xi^{-}~e^{\sigma(x)}\epsilon^{i}({\cal L}_{\phi_{0}})_{v_{i}}{\cal E}_{\mu\nu}~\partial_{+}x^{\mu}\partial_{-}x^{\nu}}-\frac{1}{2}\int{d\epsilon^{i}\wedge \star J_{i}}.
\end{equation}
Now, with $\delta S=0$ we have

\begin{equation}\label{7}
d \star J_{i}=e^{\sigma(x)}[({\cal L}_{\phi_{0}})_{v_{i}}{\cal E}_{\mu\nu}~\partial_{+}x^{\mu}\partial_{-}x^{\nu}+v_{i}\hspace{0cm}^{\lambda}(equations~of~motion)] d\xi^{-}\wedge d\xi^{+},
\end{equation}
where the equations of motion have the following form

\begin{equation}\label{8}
({\cal E}_{\lambda\nu}+{\cal E}_{\nu\lambda})~\partial_{+}\partial_{-}x^{\nu}+(\partial_{\mu}{\cal E}_{\lambda\nu}+\partial_{\nu}{\cal E}_{\mu\lambda}-\partial_{\lambda}{\cal E}_{\mu\nu}+{\cal E}_{\lambda\nu}\partial_{\mu}\sigma+{\cal E}_{\mu\lambda}\partial_{\nu}{\huge \sigma}-{\cal E}_{\mu\nu}\partial_{\lambda}\sigma)~\partial_{+}x^{\mu} \partial_{-}x^{\nu}=0.
\end{equation}
The equation \eqref{7} can be obtained by direct calculation on the relation \eqref{4}.
The solution of the equation \eqref{8} is a two-dimensional extremal surface in the target space $M$. In other words, on this extremal surface, we have $\delta S=0$ and

\begin{equation}\label{9}
d \star J_{i}=e^{\sigma(x)}[({\cal L}_{\phi_{0}})_{v_{i}}{\cal E}_{\mu\nu}~\partial_{+}x^{\mu}\partial_{-}x^{\nu}] d\xi^{-}\wedge d\xi^{+}.
\end{equation}
We see that using the above relation, the isometry symmetry is not equal to ${\cal L}_{v_{i}}({\cal E}_{\mu\nu})=0$, but we have isometry symmetry with new condition $({\cal L}_{\phi_{0}})_{v_{i}}{\cal E}_{\mu\nu}=0$. Therefore, in this new approach if $({\cal L}_{\phi_{0}})_{v_{i}}{\cal E}_{\mu\nu}=0$, then $G$ is an isometry group of M and we have conserved Noether's currents. On the other hand, we assume that the 1-forms $\star J_{i}$ are not closed and we have an ansatz that they obey the following generalized Maurer-Cartan equation \cite{IM}

\begin{equation}\label{10}
d \star J_{i}=-\frac{1}{2}e^{-\sigma}({\tilde{f}^{jk}}\hspace{0cm}_{i}-\alpha^{j}\delta^{k}
\hspace{0cm}_{i} +\alpha^{k}\delta^{j}\hspace{0cm}_{i}) \star J_{j} \wedge \star J_{k},
\end{equation}
with ${\tilde{f}^{jk}}\hspace{0cm}_{i}$ are some coefficients (we will see that they are the structure constants of  dual Lie algebra ${\tilde{\bf g}}$) and $\alpha^{i}$ are coefficients of $X_{0}=d {\tilde \sigma} \in {\bf g}$ (i.e. $X_{0}=\alpha^{i}X_{i}$ where $\{X_{i}\}$ is the basis of ${\bf g}$, and ${\tilde \sigma}$ is a function in the dual manifold $\tilde{M}$). Then, using \eqref{9} and \eqref{10} the condition of {\it Jacobi-Lie symmetry} can be formulated in the following form

\begin{equation}\label{11}
({\cal L}_{\phi_{0}})_{v_{i}}{\cal E}_{\mu\nu}=({\tilde{f}^{jk}}\hspace{0cm}_{i}-\alpha^{j}\delta^{k}
\hspace{0cm}_{i} +\alpha^{k}\delta^{j}\hspace{0cm}_{i}) v_{j}\hspace{0cm}^{\lambda} v_{k}\hspace{0cm}^{\eta} {\cal E}_{\mu\eta} {\cal E}_{\lambda\nu}.
\end{equation}
In this way, the sigma model \eqref{1} has the Jacobi-Lie symmetry under the action of Lie group $G$ on the target manifold $M$ if the background matrix ${\cal E}_{\mu\nu}$ satisfies the relation \eqref{11}. The Jacobi-Lie symmetry is the generalization of Poisson-Lie symmetry in the presence of 1-cocycles $X_{0}$ and $\phi_{0}$. In other words, when $\alpha^{i}=\beta_{i}=0$ (or $\sigma=\tilde \sigma=0$) the Jacobi-Lie symmetry \eqref{11} is converted to the well-known Poisson-Lie symmetry relation \cite{KL, K}

\begin{equation}\label{28''''}
{\cal L}_{v_{i}}{\cal E}_{\mu\nu}={\tilde{f}^{jk}}\hspace{0cm}_{i} v_{j}\hspace{0cm}^{\lambda} v_{k}\hspace{0cm}^{\eta} {\cal E}_{\mu\eta} {\cal E}_{\lambda\nu}.
\end{equation}
Now, we will consider the integrability condition for the $\phi_{0}$-Lie derivative

\begin{equation}\label{12}
({\cal L}_{\phi_{0}})_{[v_{i},v_{j}]}{\cal E}_{\mu\nu}=({\cal L}_{\phi_{0}})_{v_{i}}({\cal L}_{\phi_{0}})_{v_{j}}{\cal E}_{\mu\nu}-({\cal L}_{\phi_{0}})_{v_{j}}({\cal L}_{\phi_{0}})_{v_{i}}{\cal E}_{\mu\nu}.
\end{equation}
According to the definition of the $\phi_{0}$-Lie derivative presented in \eqref{5}, and 1-cocycle condition on $\phi_{0}$ \cite{RS4}, i.e.,

\begin{equation}\label{13}
\beta_{k}{f}_{ij}\hspace{0cm}^{k}=0,
\end{equation}
where $\beta_{k}\equiv v_{k}\hspace{0cm}^{\lambda}\partial_{\lambda}\sigma$, the integrability condition for $\phi_{0}$-Lie derivative is equivalent to the integrability condition for usual Lie derivative, i.e.

\begin{equation}\label{14}
{\cal L}_{[v_{i},v_{j}]}{\cal E}_{\mu\nu}={\cal L}_{v_{i}}{\cal L}_{v_{j}}{\cal E}_{\mu\nu}-{\cal L}_{v_{j}}{\cal L}_{v_{i}}{\cal E}_{\mu\nu}.
\end{equation}
Now, using \eqref{5} and \eqref{11}, after some computations, the integrability condition \eqref{14} gives the first condition of Jacobi-Lie bialgebra $(({\bf{g}},\phi_{0}),({\tilde{\bf g}},X_{0}))$ \cite{RS4}, as follows

$$
\hspace{-.3cm}{f}_{ij}\hspace{0cm}^k{\tilde{f}^{mn}}\hspace{0cm}_{k}-{f}_{ik}\hspace{0cm}^m{\tilde{f}^{kn}}\hspace{0cm}_{j} -\\
{f}_{ik}\hspace{0cm}^n{\tilde{f}^{mk}}\hspace{0cm}_{j}-{f}_{kj}\hspace{0cm}^m{\tilde{f}^{kn}}\hspace{0cm}_{i}-\\
{f}_{kj}\hspace{0cm}^n{\tilde{f}^{mk}}\hspace{0cm}_{i}+\beta_{i}{\tilde{f}^{mn}}\hspace{0cm}_{j}-\beta_{j}{\tilde{f}^{mn}}\hspace{0cm}_{i}+\alpha^{m}{f}_{ij}\hspace{0cm}^n-\alpha^{n}{f}_{ij}\hspace{0cm}^m\\
$$
\begin{equation}\label{15}
+(\alpha^{k}{f}_{ik}\hspace{0cm}^m-\alpha^{m}\beta_{i})\delta_{j}\hspace{0cm}^{n}
-(\alpha^{k}{f}_{jk}\hspace{0cm}^m-\alpha^{m}\beta_{j})\delta_{i}\hspace{0cm}^{n}-(\alpha^{k}{f}_{ik}\hspace{0cm}^n-\alpha^{n}\beta_{i})\delta_{j}\hspace{0cm}^{m}
+(\alpha^{k}{f}_{jk}\hspace{0cm}^n-\alpha^{n}\beta_{j})\delta_{i}\hspace{0cm}^{m}=0.
\end{equation}
In other words, the integrability condition of the $\phi_{0}$-Lie derivative together with the Jacobi-Lie symmetry \eqref{11} provide the result that the dual space ${\tilde{\bf g}}$ is a Lie algebra with structure constants ${\tilde{f}^{jk}}\hspace{0cm}_{i}$ and 1-cocycle $X_{0}$ such that the first condition of the Jacobi-Lie bialgebra $(({\bf{g}},\phi_{0}),({\tilde{\bf g}},X_{0}))$ is satisfied \cite{RS4}.

In the same way, one can consider the dual sigma model on the manifold $\tilde M$ with background matrix ${\cal \tilde E}_{\mu\nu}=\tilde G_{\mu\nu}+\tilde B_{\mu\nu}$ in the presence of a $\tilde \sigma$-function on $\tilde M$ as follows

\begin{equation}\label{16}
\tilde S=\frac{1}{2}\int{d\xi^{+}\wedge d\xi^{-}~e^{\tilde \sigma(\tilde x)}{\cal \tilde E}_{\mu\nu}(\tilde x)~ \partial_{+}\tilde x^{\mu} \partial_{-}\tilde x^{\nu}}=\frac{1}{2}\int{d\xi^{+}\wedge d\xi^{-}~\tilde L}.
\end{equation}
Now, one can consider the variation of $\tilde S$ under the action of dual Lie group $\tilde G$ on $\tilde M$ with the following transformation

\begin{equation}\label{17}
\tilde x^{\mu}\rightarrow \tilde x^{\mu}+\tilde \epsilon_{i}(\xi^{+},\xi^{-})\tilde v^{i}\hspace{0cm}^{\mu},
\end{equation}
where the left-invariant vector fields $\tilde v^{i}\hspace{0cm}^{\mu}$ correspond to the right action of the Lie group $\tilde{G}$ on the target manifold $\tilde{M}$. Therefore, similar to \eqref{6} we will have

\begin{equation}\label{18}
\delta \tilde S=\frac{1}{2}\int{d\xi^{+}\wedge d\xi^{-}~e^{\tilde \sigma(\tilde x)}\tilde \epsilon_{i}({\cal {\tilde L}}_{\tilde v^{i}}}{\cal \tilde E}_{\mu\nu}+\tilde v^{i}\hspace{0cm}^{\lambda}\tilde \partial_{\lambda}\tilde \sigma~{\cal \tilde E}_{\mu\nu})~\partial_{+}\tilde x^{\mu}\partial_{-}\tilde x^{\nu}-\frac{1}{2}\int{d\tilde \epsilon_{i}\wedge \star \tilde J^{i}}.
\end{equation}
If we consider $X_{0}=d \tilde \sigma \in \Omega^{1}(\tilde M)$ as a 1-cocycle on ${\tilde{\bf g}}$ with values in ${\bf g}$, then using $X_{0}$-Lie derivative \cite{IM}

\begin{equation}\label{19}
({\cal {\tilde L}}_{X_{0}})_{\tilde v^{i}}{\cal \tilde E}_{\mu\nu}={\cal {\tilde L}}_{\tilde v^{i}}{\cal \tilde E}_{\mu\nu}+<X_{0},\tilde v^{i}>{\cal \tilde E}_{\mu\nu},
\end{equation}
$\delta \tilde S$ can be rewritten as follows

\begin{equation}\label{20}
\delta \tilde S=\frac{1}{2}\int{d\xi^{+}\wedge d\xi^{-}~e^{\tilde \sigma(\tilde x)}\tilde \epsilon_{i}({\cal {\tilde L}}_{X_{0}})_{\tilde v^{i}}{\cal \tilde E}_{\mu\nu}~\partial_{+}\tilde x^{\mu}\partial_{-}\tilde x^{\nu}}-\frac{1}{2}\int{d\tilde \epsilon_{i}\wedge \star \tilde J^{i}},
\end{equation}
where

\begin{equation}\label{21}
\star \tilde J^{i}=e^{\tilde \sigma(\tilde x)}({\cal \tilde E}_{\mu\gamma}~\tilde v^{i}\hspace{0cm}^{\gamma}~ \partial_{+}\tilde x^{\mu}d\xi^{+}-{\cal \tilde E}_{\gamma\nu}~\tilde v^{i}\hspace{0cm}^{\gamma}~ \partial_{-}\tilde x^{\nu}d\xi^{-}).
\end{equation}
Similar to the original model, if we use the equations of motion related to the dual model $\tilde S$ and the following generalized Maurer-Cartan equation for the Noether's currents $\star \tilde J^{i}$

\begin{equation}\label{22}
d \star \tilde J^{i}=-\frac{1}{2}e^{-\tilde \sigma}({{f}_{jk}}\hspace{0cm}^{i}-\beta_{j}\delta_{k}
\hspace{0cm}^{i} +\beta_{k}\delta_{j}\hspace{0cm}^{i}) \star \tilde J^{j} \wedge \star \tilde J^{k},
\end{equation}
where ${{f}_{jk}}\hspace{0cm}^{i}$ are the structure constants of the Lie algebra ${\bf g}$ and $\beta_{i}$ are coefficients of $\phi_{0}=d {\sigma} \in {\tilde{\bf g}}$ i.e. $\phi_{0}=\beta_{i} \tilde X^{i}$, $\{\tilde X^{i}\}$ being the basis for ${\tilde{\bf g}}$ (${\sigma}$ is the function in the original model), then we have the Jacobi-Lie symmetry on the dual model as follows

\begin{equation}\label{23}
({\cal {\tilde L}}_{X_{0}})_{\tilde v^{i}}{\cal \tilde E}_{\mu\nu}=({{f}_{jk}}\hspace{0cm}^{i}-\beta_{j}\delta_{k}
\hspace{0cm}^{i} +\beta_{k}\delta_{j}\hspace{0cm}^{i}) \tilde v^{j}\hspace{0cm}^{\lambda} \tilde v^{k}\hspace{0cm}^{\eta} {\cal \tilde E}_{\mu\eta} {\cal \tilde E}_{\lambda\nu}.
\end{equation}
In the same way, the integrability condition for $({\cal {\tilde L}}_{X_{0}})_{\tilde v^{i}}$

\begin{equation}\label{24}
({\cal {\tilde L}}_{X_{0}})_{[\tilde v^{m},\tilde v^{n}]}{\cal \tilde E}_{\mu\nu}= ({\cal {\tilde L}}_{X_{0}})_{\tilde v^{m}}({\cal {\tilde L}}_{X_{0}})_{\tilde v^{n}}{\cal \tilde E}_{\mu\nu}-({\cal {\tilde{L}}}_{X_{0}})_{\tilde v^{n}}({\cal {\tilde{L}}}_{X_{0}})_{\tilde v^{m}}{\cal \tilde E}_{\mu\nu},
\end{equation}
with the 1-cocycle condition for $X_{0}$ \cite{RS4} i.e.,

\begin{equation}\label{25}
\alpha^{k}{\tilde{f}^{mn}}\hspace{0cm}_{k}=0,
\end{equation}
with $\alpha^{k}\equiv \tilde v^{k}\hspace{0cm}^{\lambda}\tilde \partial_{\lambda}\tilde \sigma$, is equivalent to the integrability condition for $({\cal {\tilde{L}}})_{\tilde v^{i}}$ as follows

\begin{equation}\label{26}
{\cal {\tilde{L}}}_{[\tilde v^{m},\tilde v^{n}]}{\cal \tilde E}_{\mu\nu}= {\cal {\tilde{L}}}_{\tilde v^{m}}{\cal {\tilde{L}}}_{\tilde v^{n}}{\cal \tilde E}_{\mu\nu}-{\cal {\tilde{L}}}_{\tilde v^{n}}{\cal {\tilde{L}}}_{\tilde v^{m}}{\cal \tilde E}_{\mu\nu}.
\end{equation}
Using \eqref{19} and \eqref{23}, after some calculation, we see that the structure constants of the Lie algebras ${\bf g}$, ${\tilde{\bf g}}$ and coefficients $\alpha^{i}$ and $\beta_{j}$ (coefficients of 1-cocycles $X_{0}$ and $\phi_{0}$, respectively) must be satisfied in the following relation

$$
\hspace{-.3cm}{f}_{ij}\hspace{0cm}^k{\tilde{f}^{mn}}\hspace{0cm}_{k}-{f}_{ik}\hspace{0cm}^m{\tilde{f}^{kn}}\hspace{0cm}_{j} -\\
{f}_{ik}\hspace{0cm}^n{\tilde{f}^{mk}}\hspace{0cm}_{j}-{f}_{kj}\hspace{0cm}^m{\tilde{f}^{kn}}\hspace{0cm}_{i}-\\
{f}_{kj}\hspace{0cm}^n{\tilde{f}^{mk}}\hspace{0cm}_{i}+\beta_{i}{\tilde{f}^{mn}}\hspace{0cm}_{j}-\beta_{j}{\tilde{f}^{mn}}\hspace{0cm}_{i}+\alpha^{m}{f}_{ij}\hspace{0cm}^n-\alpha^{n}{f}_{ij}\hspace{0cm}^m\\
$$
\begin{equation}\label{27}
+(\beta_{k}{\tilde f}^{mk}\hspace{0cm}_i-\alpha^{m}\beta_{i})\delta_{j}\hspace{0cm}^{n}
-(\beta_{k}{\tilde f}^{mk}\hspace{0cm}_j-\alpha^{m}\beta_{j})\delta_{i}\hspace{0cm}^{n}-(\beta_{k}{\tilde f}^{nk}\hspace{0cm}_i-\alpha^{n}\beta_{i})\delta_{j}\hspace{0cm}^{m}
+(\beta_{k}{\tilde f}^{nk}\hspace{0cm}_j-\alpha^{n}\beta_{j})\delta_{i}\hspace{0cm}^{m}=0.
\end{equation}
The above relation is the first condition of Jacobi-Lie bialgebra $(({\tilde{\bf g}},X_{0}),({\bf{g}},\phi_{0}))$ \cite{RS4}. Finally, with subtraction of the relations \eqref{15} and \eqref{27}, we obtain the third condition of Jacobi-Lie bialgebras $(({\bf{g}},\phi_{0}),({\tilde{\bf g}},X_{0}))$ and $(({\tilde{\bf g}},X_{0}),({\bf{g}},\phi_{0}))$, i.e.,\cite{RS4}

\begin{equation}\label{28}
\alpha^{k}{f}_{ik}\hspace{0cm}^{m}-\beta_{k}{\tilde{f}^{mk}}\hspace{0cm}_{i}=0.
\end{equation}
This condition is symmetrizing the Jacobi-Lie bialgebras with respect to $({\bf{g}},\phi_{0})$ and $({\tilde{\bf g}},X_{0})$ i.e., if $(({\bf{g}},\phi_{0}),({\tilde{\bf g}},X_{0}))$ is a Jacobi-Lie bialgebra then $(({\tilde{\bf g}},X_{0}),({\bf{g}},\phi_{0}))$ will also be a Jacobi-Lie bialgebra \cite{RS4}. Now, if $\alpha^{i}$ and $\beta_{i}$ in relations \eqref{13}, \eqref{15}, \eqref{25}, \eqref{27} and \eqref{28} are satisfied in the following relation
\begin{equation}\label{28'}
\alpha^{i}\beta_{i}=0,
\end{equation}
then, we have the second condition of the Jacobi-Lie bialgebras $(({\bf{g}},\phi_{0}),({\tilde{\bf g}},X_{0}))$ and $(({\tilde{\bf g}},X_{0}),({\bf{g}},\phi_{0}))$ \cite{RS4}. The equation \eqref{28'} can be obtained from isometry symmetry in 1-cocycles $X_{0}$ and $\phi_{0}$. In other words, if we have the following relations

\begin{equation}\label{28''}
({\cal L})_{X_{0}}{\cal E}_{\mu\nu}=0~~~~~~~,~~~~~~~({\cal {\tilde{L}}})_{\phi_{0}}{\cal \tilde E}_{\mu\nu}=0,
\end{equation}
the relation \eqref{28'} is satisfied. Therefore, if the sigma models $S$ and $\tilde S$ on $M$ and $\tilde M$ are invariant under the freely action of $G$ on $M$ and $\tilde G$ on $\tilde M$ such that we have the Jacobi-Lie symmetry, then the Lie algebras ${\bf g}$ and ${\tilde{\bf g}}$ together with 1-cocycles $X_{0}$ and $\phi_{0}$ construct the Jacobi-Lie bialgebra structure $(({\bf{g}},\phi_{0}),({\tilde{\bf g}},X_{0}))$. The Jacobi-Lie symmetry is a generalization of the Poisson-Lie symmetry \cite{KL,K} and subsequently isometry symmetry.

In the next section, we will consider this symmetry in more details for the $T$-dual sigma models on Lie groups.
\section{\bf Jacobi-Lie T-dual sigma models on Lie groups}

In this section, we want to find a simple way for solving Eqs. \eqref{11} and \eqref{23}. For this aim, we shall discuss the case in which the Lie group $G$ acts transitively and freely on the target space $M$. In this case, the manifold $M$ is the same as the Lie group $G$. For constructing the sigma models $S$ and $\tilde S$, we will consider two orthogonal linear subspaces $\epsilon^{+}$ and $\epsilon^{-}$ such that $\epsilon^{+}+\epsilon^{-}$ span the whole 2n-dimensional vector space ${\bf g}\oplus{\tilde{\bf g}}${\footnote{$dim{\bf g} = dim{\tilde{\bf g}}=n$} with the commutation relations between  $\{X_{i}\}$ and $\{\tilde{X}^{j}\}$ (the basis of the Lie algebras ${\bf g}$ and ${\tilde{\bf g}}$, respectively) as follows \cite{RS4}

$$
[X_{i},X_{j}]={f_{ij}\hspace{0cm}}^{k} X_{k}\hspace{1mm} , \hspace{1mm}[\tilde{X}^{i},\tilde{X}^{j}]={{\tilde{f}}^{ij}\hspace{0cm}}_{k} {\tilde{X}}^{k},
$$

\begin{equation}\label{29}
[X_i , \tilde{X}^j] =({\tilde{f}^{jk}}\hspace{0cm}_{i}+\frac{1}{2}\alpha^{k}\delta_{i}\hspace{0cm}^{j}-\alpha^{j}\delta_{i}\hspace{0cm}^{k})X_k +({f}_{ki}\hspace{0cm}^{j}-\frac{1}{2}\beta_{k}\delta_{i}\hspace{0cm}^{j}+\beta_{i}\delta_{k}\hspace{0cm}^{j}) \tilde{X}^k,
\end{equation}
where ${\bf g}$ and ${\tilde{\bf g}}$ are isotropic vector spaces \footnote {In general case, ${\bf g}\oplus{\tilde{\bf g}}$ is not a Lie algebra; but with additional conditions \cite{RS4}
\begin{equation}\label{32'}
{\tilde{f}^{jk}}\hspace{0cm}_{i}=\frac{1}{2}(\alpha^{j}\delta^{k}
\hspace{0cm}_{i} -\alpha^{k}\delta^{j}\hspace{0cm}_{i}),
\end{equation}
and
\begin{equation}\label{32''}
{{f}_{jk}}\hspace{0cm}^{i}=\frac{1}{2}(\beta_{j}\delta_{k}
\hspace{0cm}^{i} -\beta_{k}\delta_{j}\hspace{0cm}^{i}),
\end{equation}
${\cal D}={\bf g}\oplus{\tilde{\bf g}}$ is a Lie algebra, one can obtain these relations by imposing the Jacobi identity to ${\cal D}={\bf g}\oplus{\tilde{\bf g}}$ with \eqref{29}}, i.e.,

\begin{equation}\label{30}
<X_{i},X_{j}>=<\tilde X^{i},\tilde X^{j}>=0~,~<X_{i},\tilde X^{j}>=\delta_{i}\hspace{0cm}^{j}.
\end{equation}
It is convenient to write \cite{KL, K}

\begin{equation}\label{31}
g^{-1} \epsilon^{+} g= Span~\{X_{i}+ E^{+}\hspace{0cm}_{ij}(g)\tilde X^{j}\},~~i,j=1,...,n,
\end{equation}
\begin{equation}\label{32}
~~~~~~~~~~~~~~~~~g^{-1} \epsilon^{-} g= Span~\{X_{i}- E^{-}\hspace{0cm}_{ij}(g)\tilde X^{j}\},~~i,j=1,...,n,~~~~~~~~\forall g \in G
\end{equation}
 i.e. we have a graph $E^{\pm}: {\bf g}\rightarrow {\tilde{\bf g}}$, as follows

\begin{equation}\label{33}
E^{\pm}(g)=(a(g)\pm E^{\pm}(e) b(g))^{-1} E^{\pm}(e) d(g),
\end{equation}
such that

$$
g^{-1} X_{i} g= {a(g)}_{i}\hspace{0cm}^{j} X_{j},
$$
\begin{equation}\label{34}
g^{-1} \tilde X^{i} g= {b(g)}^{ij} X_{j} + d(g)^{i}\hspace{0cm}_{j} \tilde X^{j}.
\end{equation}
In this presentation, we have $<\epsilon^{+},\epsilon^{-}>=0$, but from \eqref{29}, we do not have the Ad-invariance on
$<X_{i},\tilde X^{j}>=\delta_{i}\hspace{0cm}^{j}$, therefore in general, unlike Poisson-Lie T-dual sigma models \cite{KL} we do not have equality  $E^{+}(g)=(E^{-}(g))^{t}$ (where $(E^{-}(g))^{t}$ is the transpose matrix of $(E^{-}(g))$). For this reason, we consider sigma model on the Lie group $G$ in the following form
\begin{equation}\label{35}
S=\frac{1}{2}\int{d\xi^{+}} \wedge d\xi^{-}~[E^{n+}(g)]_{ij} ~(g^{-1} \partial_{+} g)^{i}~(g^{-1} \partial_{-} g)^{j}.
\end{equation}
Comparing the above action with the sigma model action \eqref{1}, and using the following expansion for $(g^{-1} \partial_{+} g)^{i}$ according to the left invariant 1-forms
\begin{equation}\label{39}
(g^{-1}\partial_{\pm}g)^{i}=L_{\pm}\hspace{0cm}^{i}=L^{i}\hspace{0cm}_{\mu} \partial_{\pm} x^{\mu},
\end{equation}
and the definition of the left-invariant vector fields
\begin{equation}\label{40}
(L^{-1})_{i}\hspace{0cm}^{\mu}=v_{i}\hspace{0cm}^{\mu}~~,~~<v_{i},L^{j}>=\delta_{i}\hspace{0cm}^{j},
\end{equation}
we will have
\begin{equation}\label{42}
e^{\sigma}{\cal E}_{\mu\nu}=L^{i}\hspace{0cm}_{\mu}L^{j}\hspace{0cm}_{\nu}E^{n+}\hspace{0cm}_{ij}(g).
\end{equation}
The Jacobi-Lie symmetry \eqref{11} for ${\cal E_{\mu\nu}}$ (using the definition of $\phi_{0}$-Lie derivation and relation (\ref{42})) is converted to the following relation for $E^{n+}(g)$
\begin{equation}\label{44.1}
v_{i}\hspace{0cm}^{\lambda} \partial_{\lambda} E^{n+}_{lm}-{f}_{il}\hspace{0cm}^{k}E^{n+}_{km}-{f}_{im}\hspace{0cm}^{k}E^{n+}_{lk}-e^{-\sigma}({\tilde{f}^{jk}}\hspace{0cm}_{i}-\alpha^{j}\delta^{k}\hspace{0cm}_{i}+\alpha^{k}\delta^{j}\hspace{0cm}_{i})
E^{n+}_{lk}E^{n+}_{jm}=0.
\end{equation}
Now, we define the following relation for $E^{n+}(g)$ 
\begin{eqnarray}\label{35'}
E^{n+}(g)=(a(g)+E^{+}({e}) a^{-t}(g) \Lambda(g))^{-1} E^{+}({e}) a^{-t}(g),
\end{eqnarray}
where $\Lambda (g)$ is the Jacobi structure on the Lie group G \cite{Iglesias, RS5}. By substituting \eqref{35'} in \eqref{44.1} and using the following relation (from \eqref{34}) \cite{S3}
\begin{equation}\label{44.2}
\partial_{\mu} a(g)_{i}\hspace{0cm}^{j} = L^{l}\hspace{0cm}_{\mu} a(g)_{i}\hspace{0cm}^{k} {f}_{kl}\hspace{0cm}^{j},
\end{equation}
one can see that $\Lambda^{ij}(g)$ must be satisfied in the following condition
\begin{equation}\label{44.3}
\partial_{\mu} \Lambda^{ij} = L^{k}\hspace{0cm}_{\mu} \{e^{-\sigma}({\tilde{f}^{ij}}\hspace{0cm}_{k}-\alpha^{i}\delta^{j}\hspace{0cm}_{k}+\alpha^{j}
\delta^{i}\hspace{0cm}_{k})+{f}_{kl}\hspace{0cm}^{[i} \Lambda^{j]l}\}.
\end{equation}
For the proof of the relation \eqref{44.3}, we use the relation $\Lambda=\hat{r}-e^{-\sigma}\bar{r}$ for the coboundary Jacobi-Lie bialgebras where 2-vector $\hat{r}(\bar{r})$ on Lie group $G$ is right(left)-translation of classical $r$-matrix (see \cite{Iglesias, RS5}). Using the right and left invariant vector fields ($v^{\scriptsize r}_{i}\hspace{0cm}^{\mu}$ and $v^{\scriptsize l}_{i}\hspace{0cm}^{\mu}$), $\Lambda(g)$ is rewritten as follows
\begin{equation}\label{44.4}
\Lambda^{\mu\nu} = r^{ij} (v^{\scriptsize r}_{i}\hspace{0cm}^{\mu}v^{\scriptsize r}_{j}\hspace{0cm}^{\nu}-e^{-\sigma}v^{\scriptsize l}_{i}\hspace{0cm}^{\mu}v^{\scriptsize l}_{j}\hspace{0cm}^{\nu}).
\end{equation}
But from the $v^{\scriptsize r}_{i}\hspace{0cm}^{\mu}=a(g)_{i}\hspace{0cm}^{m} v^{\scriptsize l}_{m}\hspace{0cm}^{\mu}$ and $\Lambda^{\mu\nu}(g)=v^{\scriptsize l}_{i}\hspace{0cm}^{\mu}v^{\scriptsize l}_{j}\hspace{0cm}^{\nu} \Lambda^{ij}(g)$, after some calculations, we have the following relation
\begin{eqnarray}\label{44.5}
\Lambda^{ij}=r^{mn} a(g)_{m}\hspace{0cm}^{i} a(g)_{n}\hspace{0cm}^{j} - e^{-\sigma} r^{ij}.
\end{eqnarray}
Now, if $r^{ij}$ is independent from the group coordinates and using \eqref{44.2} and $\beta_{k}\equiv v^{l}_{k}\hspace{0cm}^{\lambda}\partial_{\lambda}\sigma$, then we have
\begin{eqnarray}\label{44.6}
\partial_{\mu} \Lambda^{ij} = L^{k}\hspace{0cm}_{\mu} (r^{mn} a(g)_{m}\hspace{0cm}^{l} a(g)_{n}\hspace{0cm}^{j}{f}_{lk}\hspace{0cm}^{i}+r^{mn} a(g)_{m}\hspace{0cm}^{i} a(g)_{n}\hspace{0cm}^{l}{f}_{lk}\hspace{0cm}^{j}+e^{-\sigma}\beta_{k} r^{ij}),
\end{eqnarray}
such that from \eqref{44.2}, \eqref{44.5} and the generalized classical Yang-Baxter equation for coboundary Jacobi-Lie bialgebras \cite{RS5}, this relation is converted to \eqref{44.3}. 

To see that the original definition of duality in a field theory \cite{FT} is satisfied for our model (i.e., the {\it equations of motion} with Jacobi-Lie symmetry of the above model are equal to the following generalized flat connection condition identity on $\tilde G$) we should choose $\tilde{\bf g}$ valued 1-forms $A_{\pm i}$ such that they satisfy in the following generalized flat connection condition

\begin{equation}\label{36}
\partial_{+}A_{-i}(g)-\partial_{-}A_{+i}(g)-e^{-\sigma}({\tilde{f}^{jk}}\hspace{0cm}_{i}-\alpha^{j}\delta^{k}\hspace{0cm}_{i} +\alpha^{k}\delta^{j}\hspace{0cm}_{i})A_{-j}(g)A_{+k}(g)=0.
\end{equation}
To this aim, we assume the ansatz that $A_{\pm i}(g)$ is the combination of 1-forms on the Lie group $G$, i.e.

\begin{equation}\label{37}
A_{\pm i}(g)=\pm E^{n\mp}\hspace{0cm}_{ij}(g)(g^{-1}\partial_{\pm}g)^{j},
\end{equation}
such that $E^{n-}(g)=(E^{n+}(g))^{t}$. Using \eqref{39}-\eqref{42}, we will have 

$$
A_{+i}(g)=e^{\sigma}v_{i}\hspace{0cm}^{\nu}{\cal E}_{\mu\nu} \partial_{+} x^{\mu},
$$
\begin{equation}\label{41}
A_{-i}(g)=- e^{\sigma}v_{i}\hspace{0cm}^{\mu}{\cal E}_{\mu\nu} \partial_{-} x^{\nu},
\end{equation}
then from \eqref{36} and \eqref{41}, the flat connection equation (generalized Maurer-Cartan equation) on $\tilde G$ is rewritten as follows

$$
-v_{i}\hspace{0cm}^{\lambda}[({\cal E}_{\lambda\nu}+{\cal E}_{\nu\lambda})~\partial_{+}\partial_{-}x^{\nu}+(\partial_{\mu}{\cal E}_{\lambda\nu}+\partial_{\nu}{\cal E}_{\mu\lambda}+{\cal E}_{\lambda\nu}\partial_{\mu}\sigma+{\cal E}_{\mu\lambda}\partial_{\nu}{\huge \sigma})~\partial_{+}x^{\mu} \partial_{-}x^{\nu}]-(\partial_{\nu}v_{i}\hspace{0cm}^{\lambda}{\cal E}_{\mu\lambda}+\partial_{\mu}v_{i}\hspace{0cm}^{\lambda}{\cal E}_{\lambda\nu})~\partial_{+}x^{\mu} \partial_{-}x^{\nu}
$$
\begin{equation}\label{43}
+({\tilde{f}^{jk}}\hspace{0cm}_{i}-\alpha^{j}\delta^{k}
\hspace{0cm}_{i} +\alpha^{k}\delta^{j}\hspace{0cm}_{i}) v_{j}\hspace{0cm}^{\lambda} v_{k}\hspace{0cm}^{\eta} {\cal E}_{\mu\eta} {\cal E}_{\lambda\nu}~\partial_{+}x^{\mu} \partial_{-}x^{\nu}=0.
\end{equation}
Finally, from the equations of motion \eqref{8} and the definition of usual Lie derivative and $\phi_{0}$-Lie derivative \eqref{5}, the above equations can be rewritten as follows

\begin{equation}\label{44}
-v_{i}\hspace{0cm}^{\lambda}~(equations~of~motion) - [({\cal L}_{\phi_{0}})_{v_{i}}{\cal E}_{\mu\nu}-({\tilde{f}^{jk}}\hspace{0cm}_{i}-\alpha^{j}\delta^{k}
\hspace{0cm}_{i} +\alpha^{k}\delta^{j}\hspace{0cm}_{i}) v_{j}\hspace{0cm}^{\lambda} v_{k}\hspace{0cm}^{\eta} {\cal E}_{\mu\eta} {\cal E}_{\lambda\nu}]~\partial_{+}x^{\mu} \partial_{-}x^{\nu}=0.
\end{equation}
Therefore, we show that the generalized flat connection condition on the Lie group $\tilde G$ is equivalent to the equations of motion of sigma model on the Lie group $G$ if we have the Jacobi-Lie symmetry for this model. Note that, in this way, we see that the sigma model \eqref{35} with \eqref{35'} has the Jacobi-Lie symmetry and the equation of motion of the original model \eqref{35} is equivalent to the flat connection in the dual model.

In the same way, one can obtain the following Jacobi-Lie symmetric $T$-dual sigma model
\begin{equation}\label{45}
\tilde S=\frac{1}{2}\int{d\xi^{+}} \wedge d\xi^{-}~\tilde E^{n+}(\tilde g)\hspace{0cm}^{ij} ~(\tilde g^{-1} \partial_{+} \tilde g)_{i}~(\tilde g^{-1} \partial_{-} \tilde g)_{j},
\end{equation}
with
\begin{equation}\label{46}
{\tilde E}^{n+}({\tilde g})=({\tilde a}({\tilde g})+{\tilde E}^{+}({\tilde e}) {\tilde a}^{-t}({\tilde g}) {\tilde \Lambda} ({\tilde g}))^{-1} {\tilde E}^{+}({ \tilde e}) {\tilde a}^{-t}({\tilde g}).
\end{equation}
For obtaining the relation between $E^{\pm}(e)$ and $\tilde E^{\pm}(\tilde e)$, it is enough to note that the graphs $E: {\bf g}\rightarrow {\tilde{\bf g}}$ and $\tilde E: {\tilde{\bf g}}\rightarrow {\bf g}$ are non-degenerate maps, so we have
\begin{equation}\label{47}
E^{\pm}(e)\tilde E^{\pm}(\tilde e)=\tilde E^{\pm}(\tilde e)E^{\pm}(e)=I.
\end{equation}
In this way, we have constructed a Jacobi-Lie T-dual sigma models \eqref{35} and \eqref{45} such that these models have the Jacobi-Lie symmetry, i.e., the equations of motion for the original model \eqref{35} are equivalent to the identities in the dual model \eqref{45} and vice versa. This is an original definition of duality in a field theory \cite{FT} and it is a generalization of Poisson-Lie T-dual sigma models on group manifolds \cite{KL}
\section{\bf An Example}
For real three dimensional Poisson-Lie groups, we do not have dual pair $(V,V.i)$ as a three dimensional real Lie bialgebra \cite{JR, HS}. Therefore, we have not Poisson-Lie T-dual sigma models for this pair. But, for the case of Jacobi-Lie groups we have the Jacobi-Lie bialgebra $((V,\phi_{0}),(V.i,X_{0}))$
where $X_{0}=-2(X_{2}+X_{3})$ and $\phi_{0}=-2\tilde{X}^{1}$ (i.e. $\alpha^{2}=\alpha^{3}=-2~,~\alpha^{1}=0$ and $\beta_{1}=-2~,~\beta_{2}=\beta_{3}=0$) and we have the nonzero commutation relations for $V$ and $V.i$ as follows \cite{RS4}
$$
[{X}_1,{X}_2]=-{X}_2,~~~~~~~~~~~~~~~~~~~~~~~~~~~~~~~~~~~~~~~~~~~[{\tilde X}^1,{\tilde X}^2]={\tilde X}^1,
$$
$$
[{X}_1,{X}_3]=-{X}_3,~~~~~~~~~~~~~~~~~~~~~~~~~~~~~~~~~~~~~~~~~~~[{\tilde X}^1,{\tilde X}^3]={\tilde X}^1,
$$
\begin{equation}\label{64}
\hspace{8.5cm}[{\tilde X}^2,{\tilde X}^3]={\tilde X}^2-{\tilde X}^3.
\end{equation}
In this section, we will consider the Jacobi-Lie T-dual sigma models on this Jacobi-Lie bialgebra. For this purpose,  we should calculate the commutation relations between $\{X_{i}\}$ and $\{\tilde X^{j}\}$. Using the relation \eqref{29}, the nonzero commutation relations are as follows
$$
[X_{1},\tilde X^{1}]=-\tilde X^{1}, ~~~~~~~~~~~~~~~~~~[X_{2},\tilde X^{2}]=X_{2},
~~~~~~~~~~~~~ [X_{3},\tilde X^{3}]=X_{3},$$
$$
[X_{1},\tilde X^{2}]=X_{1}-\tilde X^{2}, ~~~~~~~~~~~~~ [X_{2},\tilde X^{3}]=X_{2},
~~~~~~~~~~~~~~~~~~~~~~~~~~~~~~~~$$
\begin{equation}\label{65}
[X_{1},\tilde X^{3}]=X_{1}-\tilde X^{3}, ~~~~~~~~~~~~~ [X_{3},\tilde X^{2}]= X_{3}.~~~~~~~~~~~~~~~~~~~~~~~~~~~~~~~~
\end{equation}
\subsection{The original model}
For constructing sigma model \eqref{35}, we choose the following representation for $g \in G$
\begin{equation}\label{66}
g=e^{xX_{1}}e^{yX_{2}}e^{zX_{3}}.
\end{equation}
By this choice, from \eqref{34}, \eqref{44.5}, \eqref{64}, \eqref{65} and $r =
X_1 \wedge X_2+X_1 \wedge X_3+\gamma X_2 \wedge X_3$ (where $\gamma$ is a real constant) \cite{RS5}, one can obtain $a(g)$ and $\Lambda(g)$, as follows
\begin{equation}\label{67}
a(g)=\left(\begin{array}{ccc}
1 & -y & -z\\
0 & e^{x} & 0 \\
0 & 0 & e^{x}
\end{array} \right)    ~~~~~,~~~~~~  \Lambda(g)=e^{x}\left(\begin{array}{ccc}
0 & 1-e^{x} & 1-e^{x}\\
e^{x}-1 & 0 & z-y \\
e^{x}-1 & y-z & 0
\end{array} \right).
\end{equation}
Then, by choosing $E^{+}(e)$ {\footnote {$E^{-}(e)=(E^{+}(e))^{t}$}} as
\begin{equation}\label{68}
E^{+}(e)=\left(\begin{array}{ccc}
0 & 0 & 1\\
0 & 1 & 0 \\
1 & 0 & 0
\end{array} \right),
\end{equation}
and substituting \eqref{67} and \eqref{68} in \eqref{35'}, the model \eqref{35} is written as follows
\begin{equation*}
S=\frac{1}{2}\int\frac{e^{-2x}}{\lambda}{\big\{} -z(2y-z+2e^{x})\partial_{+}x\partial_{-}x-ze^{x}\partial_{+}x\partial_{-}y+[e^{x}(e^{x}-2)+(e^{x}-1)(2y-z)]\partial_{+}x\partial_{-}z
\end{equation*}
\begin{equation*}
\hspace{1.7cm}-[2z+(2y-3z)e^{x}]\partial_{+}y\partial_{-}x
+e^{x}(e^{x}-2)\partial_{+}y\partial_{-}y-(e^{x}-1)(e^{x}-2)\partial_{+}y\partial_{-}z
\end{equation*}
\vspace{0.01mm}
\begin{equation}\label{69}
\hspace{2.3cm}[z-e^{x}(e^{x}+z)]\partial_{+}z\partial_{-}x-e^{x}(e^{x}-1)\partial_{+}z\partial_{-}y+(e^{x}-1)^{2}\partial_{+}z\partial_{-}z{\big\}}~d\xi^{+}~d\xi^{-}.
\end{equation}
By identifying the above action with the sigma model action of the form \eqref{1}, the background symmetric metric $G_{\mu\nu}$ and antisymmetric tensor field $B_{\mu\nu}$ take the following forms
\begin{equation}\label{69'}
G_{\mu\nu}= \frac{1}{\lambda}\left(\begin{array}{ccc}
-z(2y-z+2e^{x}) & -z+(z-y)e^{x} & (e^{x}-1)(y-z)-e^{x}\\
-z+(z-y)e^{x} & e^{x}(e^{x}-2) & -(e^{x}-1)^{2}\\
(e^{x}-1)(y-z)-e^{x} & -(e^{x}-1)^{2} & (e^{x}-1)^{2}\\
\end{array} \right),
\end{equation}
\vspace{5mm}
\begin{equation}\label{69''}
B_{\mu\nu}=\frac{1}{\lambda}\left(\begin{array}{ccc}
0 & z+(y-2z)e^{x} & (e^{x}+y)(e^{x}-1)\\
-z-(y-2z)e^{x} & 0 & e^{x}-1\\
-(e^{x}+y)(e^{x}-1) & 1-e^{x} & 0
\end{array} \right),
\end{equation}
where $\lambda=e^{x}(e^{x}-2)+2(e^{x}-1)(y-z)$ and the function $\sigma=-2x$, which is compatible with $\sigma$ in \cite{RS5}.

\subsection{The dual model}
For constructing dual sigma model \eqref{45}, we choose the following representation for $\tilde g \in \tilde G$

\begin{equation}\label{70}
\tilde g=e^{\tilde x\tilde X_{1}}e^{\tilde y\tilde X_{2}}e^{\tilde z\tilde X_{3}}.
\end{equation}
Similar to the original model, one can obtain $\tilde a(\tilde g)$ and $\tilde \Lambda(\tilde g)$, using $\tilde r =-(1+\delta){\tilde X}^1 \wedge {\tilde X}^2+\delta{\tilde X}^1 \wedge {\tilde X}^3$ (where $\delta$ is a real constant), \cite{RS5} as follows

\begin{equation}\label{71}
\tilde a(\tilde g)=\left(\begin{array}{ccc}
e^{\tilde y+\tilde z} & 0 & 0 \\
-\tilde x e^{\tilde y+\tilde z} & e^{\tilde z} & 1-e^{\tilde z} \\
-\tilde x e^{\tilde y+\tilde z} & (1-e^{\tilde y})e^{\tilde z} & 1-(1-e^{\tilde y})e^{\tilde z}
\end{array} \right)    ~~~~~,~~~~~~  \tilde \Lambda(\tilde g)=e^{2(\tilde{y}+\tilde{z})}\left(\begin{array}{ccc}
0 & 1-e^{-\tilde{y}} & e^{-\tilde{y}}(1-e^{-\tilde{z}})\\
e^{-\tilde{y}}-1 & 0 & 0 \\
e^{-\tilde{y}}(e^{-\tilde{z}}-1) & 0 & 0
\end{array} \right),
\end{equation}
Then, by choosing $\tilde E^{+}(\tilde e)$ as
\begin{equation}\label{72}
\tilde E^{+}(\tilde e)=\left(\begin{array}{ccc}
0 & 0 & 1\\
0 & 1 & 0 \\
1 & 0 & 0
\end{array} \right),
\end{equation}
the dual sigma model will be in the following form

\begin{equation*}
\tilde S=\frac{1}{2}\int \frac{1}{\zeta}{\big\{}(e^{\tilde z}-1)\partial_{+}\tilde x\partial_{-}\tilde y+e^{\tilde z}\partial_{+}\tilde x\partial_{-}\tilde z+(e^{\tilde z}-1)(2e^{-\tilde z}-1)\partial_{+}\tilde y\partial_{-}\tilde x+e^{\tilde y+\tilde z}[2(\tilde{x}-1)(1-e^{-\tilde{z}})+e^{\tilde{y}}]\partial_{+}\tilde y\partial_{-}\tilde y
\end{equation*}
\begin{equation}\label{72'}
\hspace{1cm}+e^{\tilde y+\tilde z}[\tilde{x}(3-2e^{-\tilde{z}})+2(e^{\tilde{y}}-1)]\partial_{+}\tilde y\partial_{-}\tilde z+(2-e^{\tilde z})\partial_{+}\tilde z\partial_{-}\tilde x+\tilde{x}e^{\tilde y+\tilde z}(\partial_{+}\tilde z\partial_{-}\tilde y+2\partial_{+}\tilde z\partial_{-}\tilde z){\big\}}~d\xi^{+}~d\xi^{-}.
\end{equation}

By comparing the above model with the dual model \eqref{16}, one can read off the symmetric metric $\tilde G_{\mu\nu}$ and antisymmetric tensor field $\tilde B_{\mu\nu}$ as follows

\begin{equation}\label{69'''}
\tilde G_{\mu\nu}=\frac{1}{2-e^{\tilde z}}\left(\begin{array}{ccc}
0 & 1-e^{-\tilde z} & 1\\
1-e^{-\tilde z} & e^{\tilde y+\tilde z}(2(\tilde{x}-1)(1-e^{-\tilde{z}})+e^{\tilde{y}}) & e^{\tilde y+\tilde z}(\tilde{x}(2-e^{-\tilde{z}})+e^{\tilde{y}}-1)\\
1 & e^{\tilde y+\tilde z}(\tilde{x}(2-e^{-\tilde{z}})+e^{\tilde{y}}-1) & 2\tilde{x}e^{\tilde y+\tilde{z}}
\end{array} \right),
\end{equation}
\vspace{4mm}
\begin{equation}\label{69''''}
     \tilde B_{\mu\nu}=\frac{1}{2-e^{\tilde z}}\left(\begin{array}{ccc}
0 & 2(cosh({\tilde z})-1) & e^{\tilde{z}}-1\\
2(1-cosh({\tilde z})) & 0 & e^{\tilde y+\tilde z}(\tilde{x}(1-e^{-\tilde{z}})+e^{\tilde{y}}-1)\\
1-e^{\tilde z} & -e^{\tilde y+\tilde z}(\tilde{x}(1-e^{-\tilde{z}})+e^{\tilde{y}}-1) & 0
\end{array} \right),
\end{equation}

where $\zeta=2e^{2\tilde y+2\tilde z}-e^{2\tilde y+3\tilde z}$ and the function $\tilde \sigma=-2(\tilde y+\tilde z)$, which is compatible with $\tilde \sigma$ in \cite{RS5}.
\section{\bf Some cocluding remarks}

We have generalized the Poisson-Lie symmetry to the Jacobi-Lie symmetry and obtained the Jacobi-Lie T-duality as a generalization of Poisson-Lie T-duality. In this way, some new cases of duality are presented. Also, we have given the formalism for construction of the Jacobi-Lie T-dual sigma models on Lie groups. There are some open problems as follows: i)Investigation of WZW models which have Jacobi-Lie symmetry, ii)Construction of Jacobi-Lie T-dual sigma models with spectator fields and proving the canonical equivalence of Jacobi-Lie T-dual sigma models. This generalization is expected to provide us with better understanding the quantum features of Poisson-Lie T-duality. Some of these problems are under investigation.\\

{\bf Acknowledgments}
\smallskip

This research was supported by a research fund No. 217/D/1639 from Azarbaijan Shahid Madani University. The authors would like to thank F. Darabi for their useful comments.
\vspace{-4mm}

\end{document}